\documentclass[12pt]{iopart}
\usepackage[T1]{fontenc}
\usepackage[utf8]{inputenc}
\usepackage[numbers,square,sort&compress]{natbib}
\usepackage{changepage}
\usepackage{dcolumn}
\usepackage{xcolor}
\usepackage{subfig}
\usepackage{standalone}
\usepackage{float}
\newcolumntype{d}[1]{D{.}{.}{#1}}

\begin{document}
\title{\textit{Ab initio} study of the reactivity of ultracold RbSr$+$RbSr collisions}
\author{Marijn P. Man $^1$, Tijs Karman and Gerrit C. Groenenboom $^2$}

\address{Theoretical Chemistry \\
Institute for Molecules and Materials, Radboud University \\
Heyendaalseweg 135, 6525 AJ Nijmegen, The Netherlands}
\ead{$^1$ Marijn.Man@ru.nl, $^2$ gerritg@theochem.ru.nl}

\date{\today}
\begin{abstract}
We performed \textit{ab initio} calculations in order to assess the reactivity of ultracold  RbSr ($^2\Sigma^+$) $+$ RbSr ($^2\Sigma^+$) collisions occurring on the singlet as well as the triplet potential. At ultracold energies reactions are energetically possible if they release energy, i.e.,\ they are exoergic. The exoergicity of reactions between RbSr molecules producing diatomic molecules are known experimentally. We extend this to reactions producing triatomic molecules by calculating the binding energy of the triatomic reaction products. We find that, in addition to the formation of Rb$_2$+2Sr and Rb$_2$+Sr$_2$ in singlet collisions, also the formation of Sr$_2$Rb+Rb and Rb$_2$Sr+Sr in both singlet and triplet collisions is exoergic. Hence, the formation of these reaction products is energetically possible in ultracold collisions. For all exoergic reactions the exoergicity is larger than 1000 cm$^{-1}$. We also find barrierless qualitative reaction paths leading to the formation of singlet Rb$_2$+2Sr and both singlet and triplet Rb$_2$Sr+Sr and Sr$_2$Rb+Rb reaction products and show that a reaction path with at most a submerged barrier exists for the creation of the singlet Rb$_2$+Sr$_2$ reaction product. Because of the existence of these reactions we expect ultracold RbSr collisions to result in almost-universal loss even on the triplet potential. Our results can be contrasted with collisions between alkali-diatoms, where the formation of triatomic reaction products is endoergic, and with collisions between ultracold SrF molecules, where during triplet collisions only the spin-forbidden formation of singlet SrF$_2$ is exoergic. 
\end{abstract}


\maketitle
\section{Introduction}
Ultracold molecular gases can be used to study ultracold chemistry \cite{ospelkaus:10,hu:19,liu:20} and have applications in quantum technologies such as quantum simulation \cite{blackmore:18,carr:09}. A number of experiments have been performed that create ultracold molecular gases by combining ultracold alkali-metal atoms to form  diatoms \cite{takekoshi:14,molony:14,park:15,guo:16,rvachov:17,seeelberg:18}. In these experiments it was observed that the number of molecules in the ultracold gas decreases over time as a result of collisions between the diatoms \cite{gregory:19}. For some gases these collisions lead to reactions \cite{zuchowski:10,hu:19,ospelkaus:10,liu:21}, but in others long-lived collision complexes are formed \cite{mayle:12,mayle:13,gregory:19,gregory:20}. In current experiments most of these collision complexes are lost from the gas, but it may be possible to recover these complexes by eliminating short-range loss in repulsive box potentials \cite{gregory:20,christianen:19}, although there might be additional loss mechanisms that need to be eliminated \cite{gersema:21,bause:21}. 

Because of the low kinetic energy ultracold collisions are only energetically possible if they release energy, e.g., they are exoergic. The reactivity of ultracold collisions between different ground state (singlet) alkali-metal diatoms was studied by {\.Z}uchowski and Hutson \cite{zuchowski:10}. They found that the formation of triatomic reaction products is always endoergic, while the formation of two (different) diatomic reaction products is exoergic for some species. So for these species the formation of the diatomic reaction products is energetically possible at ultracold energies, while the formation of the triatomic reaction products is not. However, the situation is different for collisions between spin-aligned triplet alkali-metal diatoms, where the formation of triatomic reaction products is always allowed \cite{soldan:10,tomzav2:13}. Recently, ultracold collisions between two YbCu, YbAg, or YbAu molecules where studied theoretically. For these molecules some of the possible triatomic reaction products can be formed, while other triatomic reaction products cannot be formed \cite{tomza:21}. 

Ultracold molecular gases can also be formed by direct laser cooling, this has been done, e.g., to cool SrF molecules \cite{barry:14,norrgard:16,steinecker:16}. The only reaction that can occur in ultracold collisions between SrF molecules is the formation of singlet SrF$_2$ molecules \cite{meyer:11}. This could mean that spin polarizing a gas of SrF molecules might prevent these reactions from occurring \cite{meyer:11}. The existence of other loss mechanisms for the collision complex, such as photoexcitation or three-body recombination, might still mean that collisions between these molecules lead to universal loss \cite{mayle:12,mayle:13,gregory:20,christianen:19,cheuk:20}. 

Feshbach resonances between ultracold Rb and Sr atoms have been predicted \cite{zuchowskiv2:10} and observed \cite{barbe:18}. These Feshbach resonances could be used to create ultracold gases of RbSr molecules \cite{zuchowskiv2:10}. Other methods for the creation of these gases have also been proposed \cite{devolder:18,devolder:21}. These gases are particularly interesting because the ground state of the RbSr molecule is a $^2\Sigma^+$ state and has a relatively large permanent electric dipole moment, we found theoretical estimates ranging from 1.36 to 1.80 D \cite{zuchowskiv2:10,guerout:10,pototschnig:14,pototschnig:16,zuchowski:14}. These properties could be used to manipulate RbSr molecules through electric and magnetic fields \cite{pototschnig:14}. Ultracold molecules with these properties are particularly suitable for certain applications, such as the precise measurement of electromagnetic fields \cite{alyabyshev:12} and simulating quantum systems \cite{micheli:06}. 

Contrary to reactions between alkali-metal diatoms, reactions between diatoms consisting of alkali metals and alkaline-earth metals have, to our knowledge, not yet been investigated in detail. From the literature \cite{stein:08,ciamei:18,strauss:10} we know that the formation of Sr$_2$+Rb$_2$ and Rb$_2$+2Sr reaction products is energetically possible in ultracold RbSr singlet collisions, but not in triplet collisions. However, the existence of viable reaction paths and the possibility of the formation of triatomic reaction products has, to our knowledge, not been investigated. 

Here we investigate, using \textit{ab initio} calculations, the reactivity of collisions between RbSr molecules. Applying \textit{ab initio} methods to systems containing alkaline-earth atoms can be difficult, because of the importance of the $p$-orbitals. These $p$-orbitals are particularly important in beryllium. This importance is illustrated by the orbital occupancy of these atoms \cite{schmidt:10}. A single beryllium atom has an orbital occupancy of 2$s^{1.80}$2$p^{0.20}$ \cite{schmidt:10}. For strontium we find that the orbital occupancy is 5$s^{1.87}$5$p^{0.13}$. This suggests that $p$-orbitals play an important role in strontium atoms, although this role appears to be slightly smaller than the role $p$-orbitals play in beryllium atoms. Here, we calculated the orbital occupancy for strontium using a complete active space self-consistent-field (CASSCF) calculation. We use basis set A from table \ref{tab:basis}. 

This article is organised as follows. In section \ref{sec:abinitio} we describe the \textit{ab initio} calculations we use to study the reactivity. In section \ref{sec:diatom} we test these calculations by comparing our results for diatomic molecules to experimental values obtained from the literature. We then investigate, in section \ref{sec:triatom}, the potential well of the triatomic molecules, (singlet and triplet) Sr$_2$Rb and (singlet and triplet) Rb$_2$Sr. In section \ref{sec:energyDiff} we show that the formation of (singlet and triplet) Rb$_2$Sr+Sr, (singlet and triplet) Sr$_2$Rb+Rb, singlet Rb$_2$+2Sr, and singlet Rb$_2$+Sr$_2$ are exoergic. So these reactions are energetically possible at ultracold energies. We also find, as described in section \ref{sec:barrier}, barrierless qualitative reaction paths for the formation of (singlet and triplet) Sr$_2$Rb+Rb, (singlet and triplet) Rb$_2$Sr+Sr, and singlet Rb$_2$+2Sr reaction products and a qualitative reaction path with a submerged barrier for the formation of singlet Rb$_2$+Sr$_2$. In section \ref{sec:implication} we use results from the literature to discuss the possible implications of our results and in section \ref{sec:conclude} we conclude. 
\section{Method}
\subsection{\textit{Ab initio} calculations}\label{sec:abinitio}
For all \textit{ab initio} calculations performed in this paper we use the MOLPRO 2015.1 \cite{werner:11,MOLPRO} software package. The electron configurations of rubidium and strontium are given by [Kr]5$s$ and [Kr]5$s^2$, respectively. We use large-core effective core potentials (ECPs), where we describe the [Kr] core of both strontium and rubidium using the large-core ECPs and core-polarization potentials (CPPs) ECP36SDF from Fuentealba \textit{et al.\ }\cite{fuentealba:83,fuentealba:85} and von Szentpály \textit{et al.\ }\cite{vonszentpaly:82}. 

We now describe these calculations in more detail. We perform spin-unrestricted open-shell single and double excitation coupled cluster with perturbative triples [UCCSD(T)] calculations \cite{knowles:93,knowles:00,watts:93}, full configuration interaction (FCI) \cite{knowles:84,knowles:89} on the valence electrons calculations, and multi-reference configuration interaction (MRCI) calculations \cite{knowles:88,werner:88}, with Pople correction \cite{pople:09,meissner:88}, on the Sr$_2$Rb$_2$ collision complex and all its fragments. The use of the MRCI method is necessary for calculations of open-shell singlet states, since the UCCSD(T) method can only be used on closed-shell or high-spin systems. We chose to use the Pople correction over the Davidson correction since it is expected to be more accurate for systems with a small number of electrons \cite{szalay:05}. Furthermore, we use expanded versions of the basis set from Christianen \textit{et al.\ }\cite{christianen:19b}, see table \ref{tab:basis}. To compensate for the basis set superposition error (BSSE) we also applied the Boys and Bernardi counterpoise correction \cite{boys:70}, except during geometry optimization and when calculating the zero-point energy (ZPE). 

We perform an UCCSD(T) calculation, a FCI calculation, and four MRCI calculations. We denote the MRCI calculations by: MRCI-1 (used for all systems), MRCI-2 (only for two atom systems), MRCI-3 (only for three atom systems), and MRCI-4 (only for four atom systems). These four MRCI variations differ in the excitations present in the active space of the MCSCF calculation and the reference space of the MRCI calculation. We first describe the MRCI-1 variation of the calculation. The results of \textit{ab initio} methods can depend on the choice of the initial orbital guess. At every point we are interested in we apply a series of \textit{ab initio} calculations, where each \textit{ab initio} calculation uses the orbitals from the previous calculation as initial orbital guess. We start with a spin-restricted Hartree-Fock (RHF) calculation, followed by a state-averaged multi-configurational self-consistent field (MCSCF) calculation \cite{knowles:85,werner:85}, a complete active space self-consistent-field (CASSCF) calculation \cite{knowles:85,werner:85}, and a MRCI calculation. In the state-averaged MCSCF calculation we use a minimal active space. The state-averaged MCSCF calculation is only nontrivial if the system contains two rubidium atoms. In which case we perform a state-averaged calculation with equal weights over the singlet and the triplet state. For other systems we only average over one state and the active space contains only one orbital. In those cases carrying out this calculation should be comparable to carrying out an additional HF calculation and the impact of this calculation should be negligible. In the CASSCF calculation we use the active space containing all molecular orbitals correlating with either the 5$s$ or the 5$p$ atomic orbitals. For the reference space of the MRCI calculation we use the active space from the CASSCF calculation. The number of excitations to a subset of the orbitals in this active space is restricted to a maximum of 2. This subset is the set of all molecular orbitals correlating with the 5$p$ atomic orbitals. For systems containing two rubidium atoms, orbitals used for singlet states were optimized in the spin-stretched state (except during the state-averaged MCSCF calculation) for computational stability. In the supplementary material we provide a script which can be used to calculate the energy of the singlet Rb$_2$Sr state using this calculation (MRCI-1). 

In the MRCI-2, MRCI-3, and MRCI-4 calculations we add some of the remaining virtual orbitals to both the active space and the restricted space. We also change the maximum number of excitations to this restricted space. The number of virtual orbitals we add and the maximum number of excitations to the restricted space depends on the number of atoms in the system, see table \ref{tab:methods} for details. For UCCSD(T) we immediately follow the RHF calculation by an UCCSD(T) calculation. We also perform a FCI calculation. In the FCI calculation we follow the same steps as in MRCI-1, but substitute a FCI calculation for the MRCI calculation. 

We use basis set A as described in table \ref{tab:basis}. To check for convergence with respect to the one-electron basis set we perform additional calculations using basis set B and C (also from table \ref{tab:basis}). We denote the basis set used by adding /A, /B, or /C to the name of the method. We give an overview of all the calculations used in the article in table \ref{tab:methods}.

\begin{table}
\caption{The exponents of the basis sets used. All basis sets used are modifications of the basis sets from Christianen \textit{et al.\ }\cite{christianen:19b}. The basis sets are uncontracted. The basis functions corresponding to the $s$-, $p$-, and $d$-orbitals have the same exponents. We always use the same basis set for Sr and Rb. }
    \begin{indented}
\item[]
    
    \label{tab:basis}
    \begin{tabular}{@{}ll}
    \br

        \multicolumn{2}{c}{Basis set A: }\\
        Orbital&Exponents\\
        \mr
        $s$/$p$/$d$&1.0, 0.316\,227\,7, 0.1, 0.031\,622\,8\\
        $f$&0.1, 0.031\,622\,8\\
        $g$&0.08, 0.008\\
        \mr
        \multicolumn{2}{c}{Basis set B: }\\
        Orbital&Exponents\\
        \mr
        $s$/$p$/$d$&1.0, 0.562\,341\,394\,53, 0.316\,227\,844\,01, 0.177\,828\,006\,79, \\
        &0.100\,000\,049\,32, 0.056\,234\,167\,19, 0.031\,622\,8\\
        $f$&0.1, 0.056\,234\,153\,32, 0.031\,622\,8\\
        $g$&0.08, 0.025\,298\,221\,28, 0.008\\
        \mr
        \multicolumn{2}{c}{Basis set C: }\\
        Orbital&Exponents\\
        \mr
        $s$/$p$/$d$&3.162\,277, 1.0, 0.316\,227\,7, 0.1, \\
        &0.031\,622\,8, 0.01\\
        $f$&0.316\,227\,7, 0.1, 0.031\,622\,8, 0.01\\
        $g$&0.316\,227\,7, 0.08, 0.008, 0.003\,162\,28\\
         \br
    \end{tabular}
    \end{indented}
\end{table}
\begin{table}
\caption{For each \textit{ab initio} calculation used we list: the name we give to the calculation, the basis set used (see table \ref{tab:basis}), the set of active/reference orbitals used in the CASSCF/MRCI steps (Active), the subset of these orbitals for which we restrict the number of excitations in MRCI (Restricted), and the maximum number of excitations to these orbitals (Excitations). We use $\infty$ to indicate that we do not restrict the number of excitations and $5p+n$ indicates all molecular orbitals correlated with the 5$p$ atomic orbitals plus the $n$ lowest virtual orbitals. }
    \begin{indented}
\item[]
    
    \label{tab:methods}
    \begin{tabular}{@{}lllll}
    \br
          Name&Basis&Active&Restricted&Excitations\\
          \mr
         MRCI-1/A&A&5$s$, 5$p$&5$p$&2\\
         MRCI-1/B&B&5$s$, 5$p$&5$p$&2\\
         MRCI-1/C&C&5$s$, 5$p$&5$p$&2\\
         \mr
         MRCI-2/A (2 atom)&A&5$s$, $5p + 8$&-&$\infty$\\
         MRCI-3/A (3 atom)&A&5$s$, $5p + 12$&-&$\infty$\\
         MRCI-4/A (4 atom)&A&5$s$, $5p + 2$&5$p + 2$&4\\
         \mr
         FCI/A&A&-&-&-\\
         \mr
         UCCSD(T)/A&A&-&-&-\\
         UCCSD(T)/B&B&-&-&-\\
         UCCSD(T)/C&C&-&-&-\\
         \br
    \end{tabular}
    \end{indented}
\end{table}
\section{Results}
\subsection{Comparison  of diatomic properties to the literature}\label{sec:diatom}
We validate our two-atom \textit{ab initio} calculations by comparing them to experimental results for Sr$_2$, RbSr, singlet Rb$_2$, and triplet Rb$_2$ obtained from the literature. We first place the two atoms of the diatom at a distance $r$ from each other. We calculate the energy of the diatomic molecules at the MRCI-1/A or UCCSD(T)/A levels, for interatomic distances $r$ on a grid ranging from $5.7$ to $30$ $a_0$, between $5.7$ and $12$ $a_0$ the grid spacing is $0.1$ $a_0$ outside of this range the grid spacing is increased. We then fit the function
\begin{equation}
    V(r)=A+Be^{-\alpha r}+Ce^{-2\alpha r}+De^{-3\alpha r}
\end{equation}
to the points where the energy is lower than $0.6$ $E_\mathrm{min}$, where $E_\mathrm{min}$ is the lowest energy we found for the current diatom (energies are relative to the atomization energy). We use this function to calculate the potential depth ($D_e$), the equilibrium distance ($r_e$), and the vibrational constant ($\omega_e$) of the diatomic molecule. We also calculate the energy at the $r_e$ we found for MRCI-1/A using the MRCI-1/B, MRCI-1/C, and MRCI-2/A calculations and the energy from UCCSD(T)/B and UCCSD(T)/C at the $r_e$ found by UCCSD(T)/A. In table \ref{tab:diatomLiterature} we compare our results to results from the literature. 

Potential depths calculated using the MRCI-1/A calculation differ from the experimental results by less than 11\%. This difference becomes larger if we increase the basis size, for MRCI-1/B the difference is less than 15\%. If we use a larger reference space the energy does not change much, the difference between the potential depths calculated by MRCI-1/A and those calculated by MRCI-2/A is less than 1\%. Meanwhile the UCCSD(T)/A calculation underestimates the potential depth by less than 8\%, for UCCSD(T)/B calculations this difference is less than 5.1\%, and for UCCSD(T)/C it is less than 3.5\%. So the well depths found by the UCCSD(T) calculations are closer to the well depths observed in the experiments. 

We also calculate the potential depth of RbSr and Sr$_2$ using FCI/A. Energies are much closer to the MRCI-1/A results than the UCCSD(T)/A results. This suggests that the MRCI calculations more accurately account for correlation effects than the UCCSD(T) calculations. So the better agreement between the UCCSD(T) well depth and the experimental well depth could be because of cancellation of errors. We estimate the uncertainty of our calculations to be somewhat larger than the less than 15\% difference between MRCI-1/B and experiment. 

We also check for size-consistency errors. We do this by calculating the energy for some special arrangements of the atoms in the Sr$_2$Rb$_2$ complex. In these arrangements the atoms form two diatoms that are placed far apart. For both diatoms we set the interatomic distance equal to their equilibrium distance. The second diatom is placed parallel to the first diatom at a distance of $140$ $a_0$. We calculate the energies of these arrangements using the MRCI-1/A calculation. We estimate the well depth of the first diatom as the difference between the energy we found and the energy when the interatomic distance of the first diatom is $r=140$ $a_0$. The decrease in well-depth when we use this four atom system instead of the regular two atom calculation described above is listed in the rows marked ``size-consistency error'' in table \ref{tab:diatomLiterature}. The differences we find seem to be small compared to our overall error.

\begin{table}
\caption{Table of the potential depth ($D_e$), equilibrium distance ($r_e$), and principal vibrational frequency ($\omega_e$) of the different diatoms formed by $^{85}$Rb and $^{88}$Sr (unless stated otherwise). We compare \textit{ab initio} results obtained in this paper to literature values. In the rows marked ``size-consistency error'' we list how much the well-depth decreases when we use a four atom calculation instead of a two atom calculation. }
    \begin{indented}
\item[]
    
    \begin{tabular}{@{}ld{4.0}d{2.2}d{2.1}}
        \br
          &\multicolumn{1}{l}{$D_e$ (cm$^{-1}$)}&\multicolumn{1}{l}{$r_e$ ($a_0$)}&\multicolumn{1}{l}{$\omega_e$ (cm$^{-1}$)}\\
          \mr
         Sr$_2$ MRCI-1/A&1192&8.7&43\\
         \hspace{8mm} Size-consistency error (singlet)&17&\multicolumn{1}{c}{-}&\multicolumn{1}{c}{-}\\
         \hspace{8mm} Size-consistency error (triplet)&17&\multicolumn{1}{c}{-}&\multicolumn{1}{c}{-}\\
         Sr$_2$ MRCI-1/B&1237&\multicolumn{1}{c}{-}&\multicolumn{1}{c}{-}\\
         Sr$_2$ MRCI-1/C&1214&\multicolumn{1}{c}{-}&\multicolumn{1}{c}{-}\\
         Sr$_2$ MRCI-2/A&1182&\multicolumn{1}{c}{-}&\multicolumn{1}{c}{-}\\
         Sr$_2$ FCI/A&1181&\multicolumn{1}{c}{-}&\multicolumn{1}{c}{-}\\
         Sr$_2$ UCCSD(T)/A&1023&8.7&40\\
         Sr$_2$ UCCSD(T)/B&1067&\multicolumn{1}{c}{-}&\multicolumn{1}{c}{-}\\
         Sr$_2$ UCCSD(T)/C&1046&\multicolumn{1}{c}{-}&\multicolumn{1}{c}{-}\\
         Sr$_2$ (Experimental, Stein \textit{et al.\ }\cite{stein:08})&1082&8.83&40.3\\
         Sr$_2$ (\textit{Ab initio}, Skomorowski \textit{et al.\ }\cite{skomorowski:12})&1124&\multicolumn{1}{c}{-}&\multicolumn{1}{c}{-}\\
         \mr
         RbSr MRCI-1/A&1248&8.6&42\\
         \hspace{8mm} Size-consistency error (singlet)&13&\multicolumn{1}{c}{-}&\multicolumn{1}{c}{-}\\
         \hspace{8mm} Size-consistency error (triplet)&26&\multicolumn{1}{c}{-}&\multicolumn{1}{c}{-}\\
         RbSr MRCI-1/B&1290&\multicolumn{1}{c}{-}&\multicolumn{1}{c}{-}\\
         RbSr MRCI-1/C&1252&\multicolumn{1}{c}{-}&\multicolumn{1}{c}{-}\\
         RbSr MRCI-2/A&1251&\multicolumn{1}{c}{-}&\multicolumn{1}{c}{-}\\
         RbSr FCI/A&1251&\multicolumn{1}{c}{-}&\multicolumn{1}{c}{-}\\
         RbSr UCCSD(T)/A&1148&8.7&40\\
         RbSr UCCSD(T)/B&1191&\multicolumn{1}{c}{-}&\multicolumn{1}{c}{-}\\
         RbSr UCCSD(T)/C&1150&\multicolumn{1}{c}{-}&\multicolumn{1}{c}{-}\\
         RbSr (\textit{Ab initio}, Pototschnig \textit{et al.\ }\cite{pototschnig:14})&1273&8.7&42\\
         RbSr (\textit{Ab initio}, Gu{\'e}rout \textit{et al.\ }\cite{guerout:10}&1073&8.69&\multicolumn{1}{c}{-}\\
         RbSr (Experimental, Ciamei \textit{et al.\ }\cite{ciamei:18})&1158&8.68&40.32\\
         
         \mr
         Rb$_2$ X$^1\Sigma_g^+$ MRCI-1/A&4055&7.9&58\\
         \hspace{8mm} Size-consistency error&39&\multicolumn{1}{c}{-}&\multicolumn{1}{c}{-}\\
         Rb$_2$ X$^1\Sigma_g^+$ MRCI-1/B&4095&\multicolumn{1}{c}{-}&\multicolumn{1}{c}{-}\\
         Rb$_2$ X$^1\Sigma_g^+$ MRCI-1/C&4071&\multicolumn{1}{c}{-}&\multicolumn{1}{c}{-}\\
         Rb$_2$ X$^1\Sigma_g^+$ MRCI-2/A&4052&\multicolumn{1}{c}{-}&\multicolumn{1}{c}{-}\\
         Rb$_2$ X$^1\Sigma_g^+$ (Experimental, Strauss \textit{et al.\ }\cite{strauss:10})&3994&7.96&57.7\\
         $^{87}$Rb$_2$ X$^1\Sigma_g^+$ (\textit{Ab initio}, Tomza \textit{et al.\ }\cite{tomza:13})&3912&7.99&56.1\\
         \mr
         Rb$_2$ a$^3\Sigma_u^+$ MRCI-1/A&245&11.5&13\\
         \hspace{8mm} Size-consistency error&9&\multicolumn{1}{c}{-}&\multicolumn{1}{c}{-}\\
         Rb$_2$ a$^3\Sigma_u^+$ MRCI-1/B&254&\multicolumn{1}{c}{-}&\multicolumn{1}{c}{-}\\
         Rb$_2$ a$^3\Sigma_u^+$ MRCI-1/C&247&\multicolumn{1}{c}{-}&\multicolumn{1}{c}{-}\\
         Rb$_2$ a$^3\Sigma_u^+$ MRCI-2/A&245&\multicolumn{1}{c}{-}&\multicolumn{1}{c}{-}\\
         Rb$_2$ a$^3\Sigma_u^+$ UCCSD(T)/A&245&11.5&13\\
         Rb$_2$ a$^3\Sigma_u^+$ UCCSD(T)/B&254&\multicolumn{1}{c}{-}&\multicolumn{1}{c}{-}\\
         Rb$_2$ a$^3\Sigma_u^+$ UCCSD(T)/C&247&\multicolumn{1}{c}{-}&\multicolumn{1}{c}{-}\\
         Rb$_2$ a$^3\Sigma_u^+$ (Experimental, Strauss \textit{et al.\ }\cite{strauss:10})&242&11.52&13.5\\
         $^{87}$Rb$_2$ a$^3\Sigma_u^+$ (\textit{Ab initio}, Tomza \textit{et al.\ }\cite{tomza:13})&250&11.46&13.5\\
         \br
    \end{tabular}
    \end{indented}
    \label{tab:diatomLiterature}
\end{table}
\subsection{Minima of the triatomic fragments}\label{sec:triatom}
We apply geometry optimization, using MOLPRO, on the potential-energy surfaces (PESs) of all the triatomic fragments of the collision complex. In the optimization we use energies calculated using the MRCI-1/A calculation and the UCCSD(T)/A calculation. These two methods are also used to calculate the zero-point energy (ZPE). Please note that we do not use the BSSE correction in the geometry optimization and the calculation of the ZPE. We then evaluate the energy at the minima we found using the methods we described in the methods section. All triatomic complexes, except the singlet state of Rb$_2$Sr, have a global T-shaped minimum. The singlet state of Rb$_2$Sr appears to have a local T-shaped minimum and a global minimum at the linear geometry. The energy of the global minimum, as calculated by MRCI-1/A, is only 130 cm$^{-1}$ lower than the energy of the local T-shaped minimum. At our current level of accuracy we cannot be certain which of these two minima has the lowest energy. In table \ref{tab:triatomic} we describe the resulting geometries, ZPEs, energies at the minima, and the three-body part of those energies. 

The three-body part of the energy is the difference between the energy in the well, and the sum of the energies of the three constituting diatoms. For both singlet and triplet Sr$_2$Rb, the triplet state of Rb$_2$Sr, and the linear minimum of the singlet state of Rb$_2$Sr this three-body part is negative. This indicates that the three-body energy is deepening the potential well. For the T-shaped minimum of singlet Rb$_2$Sr this term is positive, indicating that the three-body energy makes the potential well less deep. Previous studies on the quartet state of triatomic alkali molecules found that in these systems the three-body energy deepened the potential well at both the global minimum and the local linear minima \cite{soldan:03,soldan:10,tomza:13b,kos:08}. For Be$_3$, Mg$_3$, and Ca$_3$ stabilizing three-body terms were also observed \cite{kaplan:00,kos:08}. 

The difference between the MRCI energies calculated using the different basis sets is smaller than 2\% and the difference between the energies calculated using MRCI-1/A and MRCI-3/A is also less than 2\%. For the MRCI calculations we also list the contribution of the Pople correction and the counterpoise correction to the energy. The magnitude of the Pople correction becomes smaller when the active space becomes larger and the magnitude of the BSSE correction becomes smaller if the basis set becomes larger. The MRCI-3/A calculation uses a very large active space and also has a very small Pople correction. The difference between the energies calculated using MRCI-1/A and UCCSD(T)/A is larger (but less than 9\%). 
\begin{table}
    \caption{Location of the minimum, energy at the minimum, and ZPE associated with the triatomic fragments of the collision complex. The locations of the minima were obtained by minimizing the energy we calculated using MRCI-1/A or UCCSD(T)/A, see table \ref{tab:methods}. Here $r_{\mathrm{A}_2}$ and $r_\mathrm{AB}$ indicate the distance between pairs of identical and different atoms respectively. Note that due to the symmetry of these minima we only have to specify two interatomic distances. The energy at the minima obtained by MRCI-1/A is evaluated using MRCI-1/A ($E_\mathrm{1,A}$), MRCI-1/B ($E_\mathrm{1,B}$), MRCI-1/C ($E_\mathrm{1,C}$), and MRCI-3/A ($E_\mathrm{3,A}$). The energies at the minima obtained by UCCSD(T)/A are evaluated using UCCSD(T)/A ($E_\mathrm{A}$), UCCSD(T)/B ($E_\mathrm{B}$), and UCCSD(T)/C ($E_\mathrm{C}$). The energy at stage 4 of our qualitative reaction paths, see section \ref{sec:barrier}, is given in column $E_\mathrm{path}$. We indicate the contribution of the Pople correction and the contribution of the counterpoise correction to the energy by Pople or CP respectively. The three-body energies for MRCI-1/A and UCCSD(T)/A are indicated by $E^\mathrm{3b}_\mathrm{2,A}$ and $E^\mathrm{3b}_\mathrm{A}$ respectively. Values for the ZPE are calculated using MRCI-1/A or UCCSD(T)/A. }
    \makebox[\linewidth]{
    
    \begin{tabular}{@{}ld{2.2}d{2.2}d{5.0}d{5.0}d{5.0}d{5.0}d{5.0}d{5.0}d{5.0}}
    \br
    \multicolumn{8}{c}{MRCI}\\
          & \multicolumn{1}{l}{$r_{\mathrm{A}_2}$}&\multicolumn{1}{l}{$r_{\mathrm{AB}}$}&\multicolumn{1}{l}{$E_\mathrm{1,A}$}&\multicolumn{1}{l}{$E_\mathrm{1,B}$}&\multicolumn{1}{l}{$E_\mathrm{1,C}$ }&\multicolumn{1}{l}{$E_\mathrm{4,A}$}&\multicolumn{1}{l}{$E_\mathrm{path}$}&\multicolumn{1}{l}{$E^\mathrm{3b}_\mathrm{1,A}$ }&\multicolumn{1}{l}{ZPE}\\
          & \multicolumn{1}{l}{($a_0$)}&\multicolumn{1}{l}{($a_0$)}&\multicolumn{1}{l}{(cm$^{-1}$)}&\multicolumn{1}{l}{(cm$^{-1}$)}&\multicolumn{1}{l}{(cm$^{-1}$)}&\multicolumn{1}{l}{(cm$^{-1}$)}&\multicolumn{1}{l}{(cm$^{-1}$)}&\multicolumn{1}{l}{(cm$^{-1}$)}&\multicolumn{1}{l}{(cm$^{-1}$)}\\
        \mr
        Sr$_2$Rb&8.03&8.39&-4953&-5020&-4957&-5012&-5021&-1482&78\\
        \hspace{4mm}\emph{Pople}&&&\emph{-161}&\emph{-164}&\emph{-162}&\emph{-4}&\emph{-98}&&\\
        \hspace{4mm}\emph{CP}&&&\emph{340}&\emph{132}&\emph{255}&\emph{340}&\emph{301}&&\\
        Rb$_2$Sr&&&&&&&\\
        \hspace{4mm}Triplet & 9.41&8.28&-4136&-4193&-4131&-4156&-4146&-1791&69\\
        \hspace{8mm}\emph{Pople}&&&\emph{0}&\emph{0}&\emph{0}&\emph{0}&\emph{-71}&&\\
        \hspace{8mm}\emph{CP}&&&\emph{254}&\emph{191}&\emph{144}&\emph{254}&\emph{218}&&\\
        \hspace{4mm}Singlet&&&&&&&&\\
        \hspace{8mm}T-shaped & 7.90&9.56&-5473&-5522&-5466&-5428&-5452&720&51\\
        \hspace{12mm}\emph{Pople}&&&\emph{-81}&\emph{-84}&\emph{-85}&\emph{-1}&\emph{-209}&&\\
        \hspace{12mm}\emph{CP}&&&\emph{249}&\emph{191}&\emph{133}&\emph{249}&\emph{253}&&\\
        \hspace{8mm}Linear &16.05&8.02&-5603&-5683&-5625&-5577&-5555&-3299&82\\
        \hspace{12mm}\emph{Pople}&&&\emph{-46}&\emph{-47}&\emph{-100}&\emph{-1}&\emph{-136}\\
        \hspace{12mm}\emph{CP}&&&\emph{252}&\emph{197}&\emph{156}&\emph{252}&\emph{256}&&\\
        \mr
        \multicolumn{8}{c}{UCCSD(T)}\\
         & \multicolumn{1}{l}{$r_{\mathrm{A}_2}$ }&\multicolumn{1}{l}{$r_{\mathrm{AB}}$}&\multicolumn{1}{l}{$E_\mathrm{A}$ }&\multicolumn{1}{l}{$E_\mathrm{B}$ }&\multicolumn{1}{l}{$E_\mathrm{C}$ }&\multicolumn{1}{l}{$E^\mathrm{3b}_\mathrm{A}$ }&\multicolumn{1}{l}{ZPE }&\\
         & \multicolumn{1}{l}{($a_0$)}&\multicolumn{1}{l}{($a_0$)}&\multicolumn{1}{l}{(cm$^{-1}$)}&\multicolumn{1}{l}{(cm$^{-1}$)}&\multicolumn{1}{l}{(cm$^{-1}$)}&\multicolumn{1}{l}{(cm$^{-1}$)}&\multicolumn{1}{l}{(cm$^{-1}$)}&&\\
        \mr
        
        Sr$_2$Rb&8.04&8.38&-4567&-4644&-4583&-1514&77\\
        Rb$_2$Sr&&&&&&&&\\
        \hspace{4mm}Triplet &9.42&8.29&-3884&-3946&-3886&-1745&69\\
        \br
    \end{tabular}
    }
    \label{tab:triatomic}
\end{table}
\subsection{Reaction energies}\label{sec:energyDiff}
\begin{figure}
    \centering
    \includegraphics[width=0.7\textwidth]{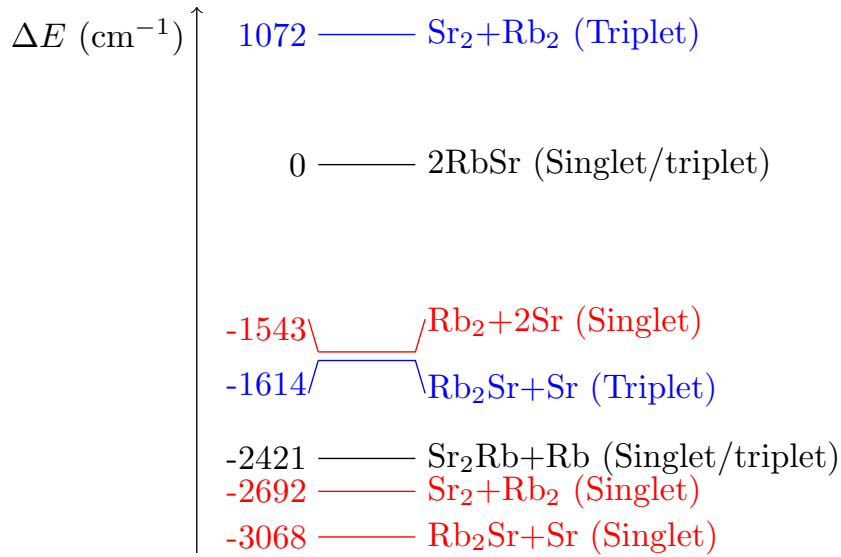}
    \caption{Reaction energy ($\Delta E$) for the reactions that can occur during RbSr+RbSr collisions. Values are calculated using the MRCI-1/A calculation. Nonpositive values of $\Delta E$ indicate that the reaction is energetically possible. Red, blue, and black lines indicate reactions found in singlet, triplet, or both types of collisions, respectively. }
    \label{fig:reactions}
\end{figure}

In the table \ref{tab:Ediff} we list the reaction energies (including the contribution from the ZPE) of all possible reactions between two RbSr molecules. The well depth and the ZPE are both calculated using the same calculation either MRCI-1/A or UCCSD(T)/A. For diatoms these properties can be found in table \ref{tab:diatomLiterature} and for triatomic molecules in table \ref{tab:triatomic}. In the ultracold limit only reactions with $\Delta E\leq 0$ are energetically possible. The table shows that for the singlet state the production of Rb$_2$Sr+Sr, Sr$_2$Rb+Rb, Rb$_2$+2Sr, or Rb$_2$+Sr$_2$ is energetically possible, while for the triplet state only the production of Rb$_2$Sr+Sr or Sr$_2$Rb+Rb is energetically possible. This is further illustrated in figure \ref{fig:reactions}. 
\begin{table}
\caption{Reaction energy ($\Delta E$) for the singlet and triplet versions of the reactions that can occur during RbSr+RbSr collisions. Nonpositive values indicate that the reaction is energetically possible. }
    \begin{indented}
\item[]
    
    \begin{tabular}{@{}lrrr}
        \br
        &\multicolumn{1}{l}{MRCI-1/A}&\multicolumn{1}{l}{MRCI-1/A}&\multicolumn{1}{l}{CCSD(T)/A}\\
        &\multicolumn{1}{l}{ Singlet}&\multicolumn{1}{l}{Triplet}&\multicolumn{1}{l}{Triplet}\\
         Product&\multicolumn{1}{l}{$\Delta E$ (cm$^{-1}$)}&\multicolumn{1}{l}{$\Delta E$ (cm$^{-1}$)}&\multicolumn{1}{l}{$\Delta E$ (cm$^{-1}$)}\\
          
         \mr
        Rb$_2$+2Sr&-1543&2222&2023\\
        Sr$_2$+2Rb&1304&1304&1272\\
        Rb$_2$+Sr$_2$&-2692&1072&1041\\
        Rb$_2$Sr+Sr&-2968&-1614&-1561\\
        Rb$_2$Sr (linear)+Sr&-3068& \multicolumn{1}{r}{-}& \multicolumn{1}{r}{-}\\
        Sr$_2$Rb+Rb&-2421&-2421&-2235\\
        \br
    \end{tabular}
    \end{indented}
    \label{tab:Ediff}
\end{table}
\subsection{Existence of reaction barriers}\label{sec:barrier}
\begin{figure}
    \centering
    \includegraphics[width=\textwidth]{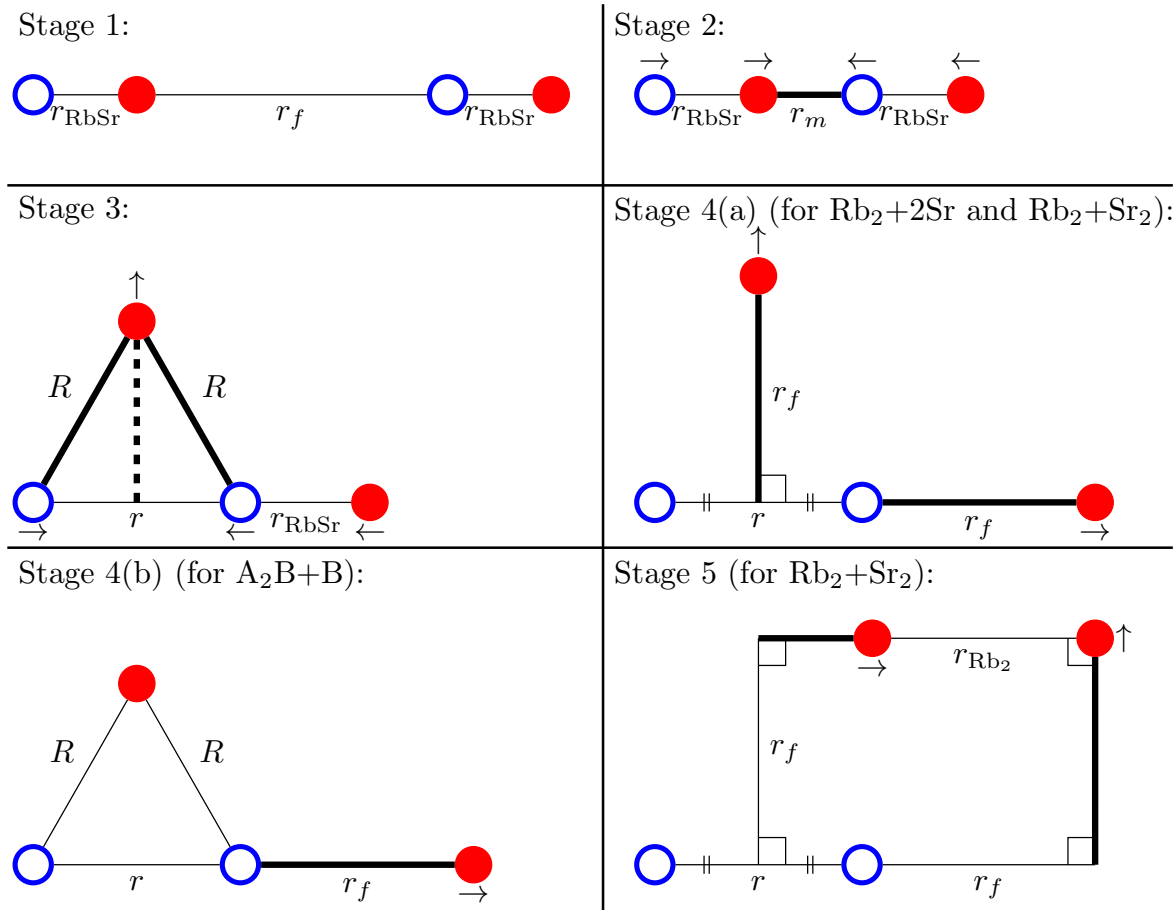}
    \caption{Diagrams of the movement of the atoms along the qualitative reaction paths. The frames display the positions of the atoms at the stages of the reaction path. We get the positions of the atoms as we move along the reaction coordinate by interpolating between these points. The arrows indicate the movement of the atoms before reaching the stage and the bold lines indicate coordinates over which we interpolate. Stage 1 represents the situation before the reaction occurs and stage 4 represents the situation after the reaction. The molecules in stage 1 are oriented so that their dipoles are aligned. The blue circles represent Rb if (singlet or triplet) Rb$_2$Sr+Sr, singlet Rb$_2$+2Sr, or singlet Rb$_2$+Sr$_2$ is created. If (singlet or triplet) Sr$_2$Rb+Rb is created they represent Sr. The red circles represent the other type of atom. }
    \label{fig:traj_steps}
\end{figure}

In order to determine whether there are barriers that suppress the exoergic reactions we devised qualitative reaction paths from the reactants to the (singlet and triplet) Rb$_2$Sr+Sr, (singlet and triplet) Sr$_2$Rb+Rb, singlet Rb$_2$+2Sr, and singlet Rb$_2$+Sr$_2$ product states. We describe these reaction paths by specifying the arrangement of the atoms at four (or five for singlet Rb$_2$+Sr$_2$) important stages, shown in figure \ref{fig:traj_steps}. We determine the arrangement of the atoms as we move along the reaction coordinate by applying linear interpolation to the coordinates describing the arrangement at these stages. The coordinates we use for interpolation are indicated by bold lines in figure \ref{fig:traj_steps}. We now define the parameters shown in figure \ref{fig:traj_steps}. We take $r_f=25$ $a_0$ for all collisions, $r_m=17$ $a_0$ for singlet collisions, and $r_m=18$ $a_0$ for triplet collisions. The coordinates $r_\mathrm{RbSr}$, $r$, and $R$ (except for $R$ in the singlet Rb$_2$+2Sr and singlet Rb$_2$+Sr$_2$ reactions) are uniquely determined by two requirements: (1) at stage 1 we have two diatoms at their equilibrium distance separated by a large distance ($r_f=25$ $a_0$) and (2) at stage 4 we have one molecule at the bottom of its potential well and one or two separate atoms. The relevant values of these parameters can then be determined from tables \ref{tab:diatomLiterature} and \ref{tab:triatomic}. For the singlet Rb$_2$+2Sr and singlet Rb$_2$+Sr$_2$ reactions we set $R\approx 9.7$ $a_0$. 

\begin{figure}
    \centering

    \subfloat[ ]{\includegraphics[width=0.8\textwidth]{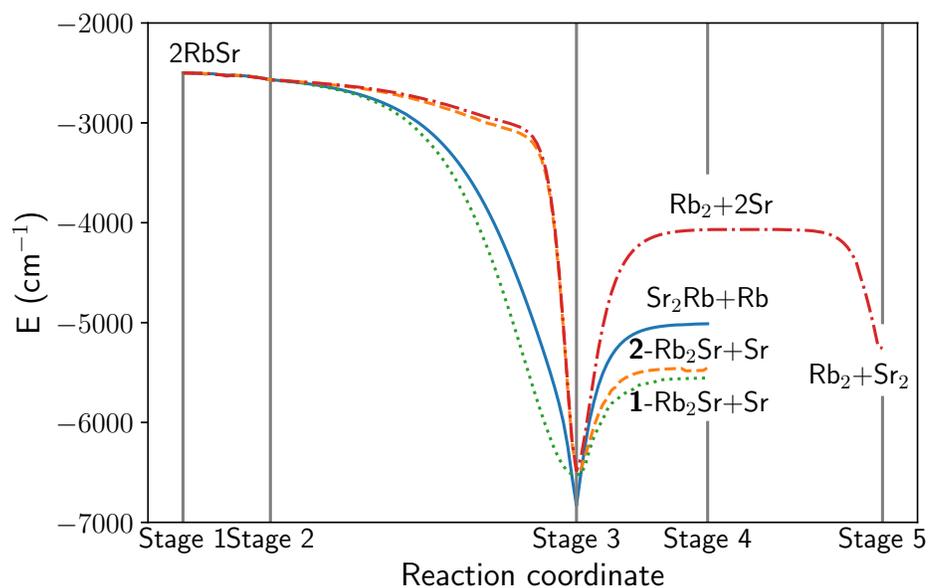}}\\
    \subfloat[ ]{\includegraphics[width=0.8\textwidth]{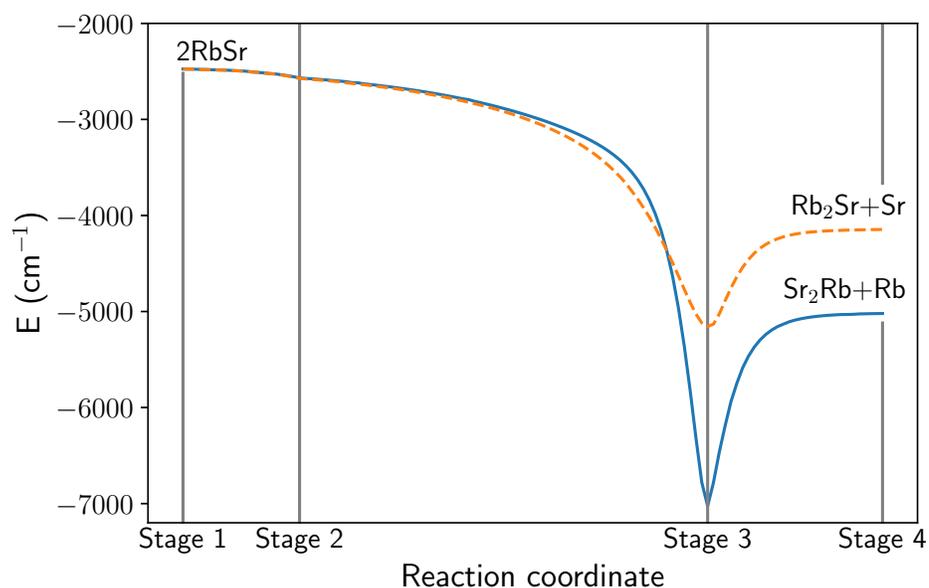}}
    
    \caption{Plots of the energy as the system moves along several qualitative reaction paths for RbSr+RbSr collisions, for reactions on the singlet (a) and triplet (b) potentials. On the right side of each line the reaction products are shown. Except for singlet Rb$_2$+2Sr, which is shown at stage 4 of the reaction path for the creation of singlet Rb$_2$+Sr$_2$. The gray lines indicate the stages between which we interpolate. We see no barriers that prevent these reactions from occurring. All energies are relative to the atomization energy. \textbf{1}-Rb$_2$Sr and \textbf{2}-Rb$_2$Sr indicate the T-shaped and the linear minima respectively. }
    \label{fig:traj_E_sin}
\end{figure}

We plot the energy, calculated using the MRCI-1/A calculation, along these approximate reaction paths for the singlet and triplet potentials in figures \ref{fig:traj_E_sin} (a) and \ref{fig:traj_E_sin} (b), respectively. The reaction paths we found are relatively simple and (with the exception of the reaction path for singlet Rb$_2$+Sr$_2$) do not contain barriers that could prevent reactions from occurring. There might be another reaction path with a lower energy. So for the creation of singlet Rb$_2$+Sr$_2$ a reaction path with at most a submerged barrier exists. When determining the possible reaction products it is important to determine which reaction paths (with at most a submerged barriers) exist, further information about the reaction paths is less important. We expect that we can approximately describe RbSr collisions by assuming they behave classically, with chaotic dynamics \cite{croft:14}. We expect this classical chaotic system to eventually reach one of the product states, either by moving along a trajectory similar to the qualitative reaction path we found or through a different route. 

In table \ref{tab:Epath} we compare the energy at the lowest point in these reaction paths (calculated using MRCI-1/A) to energies calculated using MRCI-1/B, and MRCI-4/A. We find differences that are smaller than 2\%. We also compare the energies calculated at stage 4 of the reaction paths corresponding to the formation of the triatomic molecules ($E_\mathrm{path}$) to the $E_\mathrm{1,A}$ triatomic energies we calculated before, see table \ref{tab:triatomic}. We find that the differences between $E_\mathrm{1,A}$ and $E_\mathrm{path}$ are less than 2\%. So the accuracy of the energies we calculated along our reaction paths seems to be comparable to the accuracy of the well depths we found for the triatomic molecules. 

\begin{table}
\caption{Energy at the deepest points in our qualitative reaction paths. We compare values calculated using MRCI-1/A to values calculated using MRCI-1/B, or MRCI-4/A. All energies are relative to the atomization energy. The contribution of the Pople and counterpoise correction to the total energy is indicated by Pople and CP respectively. Due to convergence problems we could not extract the energy at the deepest point for the MRCI-1/C calculation, there were also some convergence issues for the MRCI-4/A calculations. }
    \begin{indented}
\item[]
    
    \begin{tabular}{@{}lrrrr}
        \br
        \multicolumn{3}{c}{Singlet}\\
        &\multicolumn{2}{c}{Deepest point}\\
        Product of path&\multicolumn{1}{l}{MRCI-1/A }&\multicolumn{1}{l}{MRCI-1/B}&\multicolumn{1}{l}{MRCI-4/A}\\
        &\multicolumn{1}{l}{ (cm$^{-1}$)}&\multicolumn{1}{l}{(cm$^{-1}$)}&\multicolumn{1}{l}{(cm$^{-1}$)}\\
        \mr
        Rb$_2$Sr (T-shaped)+Sr&-6495&-6542&-6482\\
        \hspace{12mm}\emph{Pople}&\emph{-384}&\emph{-345}&\emph{-172}\\
        \hspace{12mm}\emph{CP}&\emph{424}&\emph{237}&\emph{424}\\
        Rb$_2$Sr (linear)+Sr&-6521&-6635&-6555\\
        \hspace{12mm}\emph{Pople}&\emph{-273}&\emph{-284}&\emph{-127}\\
        \hspace{12mm}\emph{CP}&\emph{414}&\emph{235}&\emph{414}\\
        Sr$_2$Rb+Rb&-6833&-6933&-6874\\
        \hspace{12mm}\emph{Pople}&\emph{-347}&\emph{-327}&\emph{-123}\\
        \hspace{12mm}\emph{CP}&\emph{439}&\emph{229}&\emph{439}\\
        Rb$_2$+Sr$_2$&-6482&-6531&-6474\\
        \hspace{12mm}\emph{Pople}&\emph{-372}&\emph{-338}&\emph{-172}\\
        \hspace{12mm}\emph{CP}&\emph{422}&\emph{236}&\emph{422}\\
        \mr
         \multicolumn{3}{c}{Triplet}\\
         &\multicolumn{2}{c}{Deepest point}\\
         Product of path&\multicolumn{1}{l}{MRCI-1/A }&\multicolumn{1}{l}{MRCI-1/B}&\multicolumn{1}{l}{MRCI-4/A}\\
         &\multicolumn{1}{l}{(cm$^{-1}$)}&\multicolumn{1}{l}{(cm$^{-1}$)}&\multicolumn{1}{l}{(cm$^{-1}$)}\\
         \mr
        Rb$_2$Sr+Sr&-5155&-5181&-5233\\
        \hspace{12mm}\emph{Pople}&\emph{-130}&\emph{-131}&\emph{-55}\\
        \hspace{12mm}\emph{CP}&\emph{381}&\emph{235}&\emph{381}\\
        Sr$_2$Rb+Rb&-7026&-7080&-7124\\
        \hspace{12mm}\emph{Pople}&\emph{-183}&\emph{-173}&\emph{-67}\\
        \hspace{12mm}\emph{CP}&\emph{389}&\emph{229}&\emph{389}\\
        \br
    \end{tabular}
    
    \label{tab:Epath}
    \end{indented}
\end{table}
\section{Discussion}\label{sec:implication}
Mayle \textit{et al.\ }\cite{mayle:12,mayle:13} propose that the rate at which an ultracold collision complex decays into some decay product can be estimated by the Rice-Ramsperger-Kassel-Marcus (RRKM) rate
\begin{equation}
    k=\frac{N}{h\rho}, 
\end{equation}
where $N$ is the number of exit channels corresponding to this decay product and $\rho$ the density of states of the collision complex \cite{croft:14}. So for collisions resulting in collision complexes the probability of elastic scattering is
\begin{equation}
    p_\mathrm{el}=\frac{N_\mathrm{el}}{N_\mathrm{r}}, 
\end{equation}
where $N_\mathrm{el}$ is the number of elastic exit channels and $N_\mathrm{r}$ is the number of reactive exit channels \cite{mayle:13}. In ultracold collisions $N_\mathrm{el}$ is equal to 1 \cite{mayle:13}. In contrast $N_\mathrm{r}$ can be very large, since the energy released during the reaction can be used to excite the reaction products to higher rovibrational states \cite{mayle:13}. For KRb+KRb collisions $N_\mathrm{r}=711$ (ignoring nuclear spin) \cite{liu:21}, while $\Delta E_\mathrm{KRb}$ is only approximately $9.77$ cm$^{-1}$ \cite{liu:21}. For every reaction product of the RbSr+RbSr collisions $\Delta E$ is more than 1000 cm$^{-1}$, so we expect an even larger number of exit channels. This implies that $p_\mathrm{el}\approx 0$ and almost all complex-forming collisions lead to reactions, so we would observe almost universal loss of the molecules from the gas \cite{mayle:13}. For $p$-wave collisions of KRb molecules the observed rate shows good agreement with the universal loss rate, while for $s$-wave collisions of KRb these rates agree within roughly a factor of 2 \cite{tao:10,christianen:21}. Based on these considerations we expect that the value of the cross section for RbSr+RbSr collisions would be comparable to the value predicted by universal loss. 
\section{Conclusion}\label{sec:conclude}
Using \textit{ab initio} calculations we have shown that RbSr+RbSr collisions on both the singlet and the triplet potential can lead to exoergic reactions. Furthermore, we have shown that qualitative barrierless reaction paths for the formation of (singlet and triplet) Rb$_2$Sr+Sr, (singlet and triplet) Sr$_2$Rb+Rb, and singlet Rb$_2$+2Sr exist. We also show that for the formation of singlet Rb$_2$+Sr$_2$ a reaction path with at most a submerged barrier exists. 

These results indicate that, contrary to the results obtained for ground state alkali-metal diatoms \cite{zuchowski:10}, collisions between ultracold RbSr molecules can lead to the formation of triatomic reaction products. Furthermore, contrary to the results for SrF \cite{meyer:11}, these reactions can occur during triplet collisions without the need for spin-forbidden transitions. During collisions between other ultracold diatoms triatomic reactions can occur: in YbCu, YbAg, or YbAu some (but not all) triatomic reaction products can be produced \cite{tomza:21} and during collisions between spin-aligned triplet alkali-metal diatoms triatomic reaction products can also be formed \cite{soldan:10,tomzav2:13}. 

We therefore expect gases of RbSr to have loss rates comparable to the loss rate predicted by universal loss. In experiments these reactions could potentially still be prevented by implementing shielding using DC electric fields \cite{quemener:16,matsuda:20} or microwave radiation \cite{karman:18,karman:19,anderegg:21,lassabliere:18}. 
\ack
We want to thank Mateusz Borkowski for reading a draft of this paper and providing useful suggestions for improvement. Funding for this work was provided by the Dutch Research Council (NWO) (Vrije Programma 680.92.18.05). 
\section*{References}
\renewcommand{\bibsection}{}
\bibliography{main} 
\end{document}